%% file: main.tex
\documentclass{egpubl}
\ifpdf
  \pdfpagewidth=\paperwidth
  \pdfpageheight=\paperheight
\else
  \AtBeginDvi{}
\fi
\usepackage{icat-egve2025}

 \WsPaperJoint      %

\usepackage[T1]{fontenc}
\usepackage{dfadobe}

\usepackage{cite}  %
\BibtexOrBiblatex
\electronicVersion
\PrintedOrElectronic
\usepackage{graphicx}

\usepackage{egweblnk}

\title[Preliminary Demonstration of Aerial 2D Projection via CT-Based Ray Superposition in Fog]%
      {Preliminary Demonstration of Aerial 2D Projection via CT-Based Ray Superposition in Fog}

\author[M. Nishizawa, M. Yamada, G. Sakai \& T. Nakata]
{\parbox{\textwidth}{\centering
       M. Nishizawa$^{1}$\orcid{0000-0002-8712-5820},
       M. Yamada$^{1}$,
       G. Sakai$^{1}$
       and T. Nakata$^{1}$
    }
    \\
{\parbox{\textwidth}{\centering $^1$Toyama Prefectural University, Japan\\
      }}
}

\begin{document}

\maketitle
\begin{abstract}
Volumetric and aerial displays have been pursued as a means to present three-dimensional images viewable from any direction without head-mounted devices or glasses. However, existing approaches often suffer from limitations such as restricted viewing zones, narrow display volumes, or dependence on large screens or mechanically moving components. To address these challenges, we propose a new approach based on computed tomography (CT), which applies CT principles to projector rays visualized in fog. In this method, images are reconstructed in mid-air by superimposing rays projected from multiple angles, eliminating the need for physical projection surfaces. In this paper, we report a preliminary demonstration of this approach. We built a prototype with ten scanning laser projectors arranged in a circular configuration and tested it in a fog-filled environment. The system successfully produced faint yet recognizable two-dimensional aerial images consistent with the intended shapes. This experiment confirms the feasibility of ray-superimposed back-projection in real space, not just in simulation. This preliminary result represents an initial step toward scalable, screenless, and multi-viewable volumetric displays.



\end{abstract}
\section{Introduction}
Autostereoscopic three-dimensional (3D) displays, which do not require head-mounted devices or special glasses, enable multiple viewers to observe 3D content simultaneously. This feature makes them attractive for applications such as public signage, stage performances, and education. Two major approaches have been widely studied: light-field displays \cite{Kara23-LFD} and volumetric displays \cite{Hahn23-Volumetric}. The former provide binocular disparity by directing different images to the two eyes, but the viewing zone is limited to the area in front of the display. The latter represent 3D objects as point clouds formed by light emission or scattering in space, allowing observation from any direction around 360\textdegree{}.

Most existing volumetric approaches, however, require physical screens or moving parts in the display volume, which limits scalability and installation flexibility. Techniques that generate luminous points directly in mid-air have also been proposed, but they still suffer from very restricted display regions and/or limited expressive capability.

In this work, we propose a new volumetric display method that applies the back-projection principle of X-ray computed tomography (CT) to superimposed projector beams. By reconstructing aerial images through this principle, the method has the potential to display complex 3D shapes including concavities, as well as planar patterns with intensity gradations, directly in mid-air. This paper describes the proposed method and preliminary experiments using a proof-of-concept prototype with ten projectors.

\section{Related Works}

Prior research can be broadly classified into two categories. The first employs physical screens or structures. Representative examples include rotating-screen displays \cite{Langhans02-FELIX}, LED arrays arranged in a 3D matrix to represent voxels \cite{Gothwal18-Cube}, and projection into scattering-material blocks \cite{Nayar07-PassiveScatter}. These approaches provide a certain level of display quality, but they require physical structures and moving parts, which pose challenges for scalability and installation flexibility. Moreover, the presence of physical barriers between the display and the user makes direct interaction with the rendered content difficult.

The second approach generates luminous points directly in mid-air. Since no physical structure is required in the display volume, these methods allow users to directly perceive and interact with the display volume, which makes them promising for a wide range of applications. For example, Ochiai et al. \cite{Ochiai16-FairyLights} proposed a method that employs femtosecond lasers to induce plasma emission in mid-air, although the drawing area is extremely limited due to point-by-point scanning. Smalley et al. \cite{Smalley18-Photophoretic} realized high-resolution aerial imagery by optically trapping and scanning white particles, but this approach is also restricted in frame rate and display volume. More recently, drone shows \cite{Sin23-Dronetic}, which draw point clouds in the sky by arranging drones as voxels, have become popular in media art. However, this approach faces the opposite difficulty: it is hard to downscale, and rapid deformation or movement of the point cloud is not feasible. Other examples are projection onto a curtain of fog \cite{Tokuda17-MistForm,Lam15-VolumetricFog}, but these methods are closer to 2D projection rather than a volumetric display in the strict sense.

In contrast, there are also approaches that use fog but visualize intersections of projector beams as luminous points within the fog particles, treating them as voxels. RayGraphy \cite{Yamada20-RayGraphy} and Light of Birth \cite{Asai16-LightOfBirth} are representative examples, demonstrating aerial imagery without physical screens. These methods differ fundamentally from fog-screen projection, as they enable direct volumetric representation. They also offer high scalability---limited only by the extent of the fog and projector reach---and maintain high frame rates because scanning of the screen or other equipment is unnecessary. However, since the technique projects contours or silhouettes of 3D shapes from the projector's viewpoint, the imagery is restricted to monochromatic silhouettes, making it difficult to depict concave geometries or textured surfaces with gradation. Furthermore, a solid theoretical framework for these approaches has not yet been established.

The primary contribution of this work is to provide a theoretical framework, based on CT back-projection, to a line of prior demonstrations and media-art installations that presented aerial images by superimposed light beams. By designing the projection patterns according to CT back-projection, our method is, in principle, capable of generating three-dimensional point clouds of almost arbitrary shapes, colors and brightness, provided that a sufficient number of projectors are available. Beyond this theoretical grounding, we also present a small-scale implementation that demonstrates the feasibility of this approach in practice, providing a basis for more general and higher-resolution volumetric displays.

\section{Method}

We apply the back-projection method of CT to projector rays visualized in fog, in order to generate aerial images. In this paper, we focus on the case of projecting two-dimensional images onto a horizontal plane in mid-air.

\subsection{Projection data calculation}

\begin{figure}[tb]
  \centering
  \includegraphics[width=.8\linewidth]{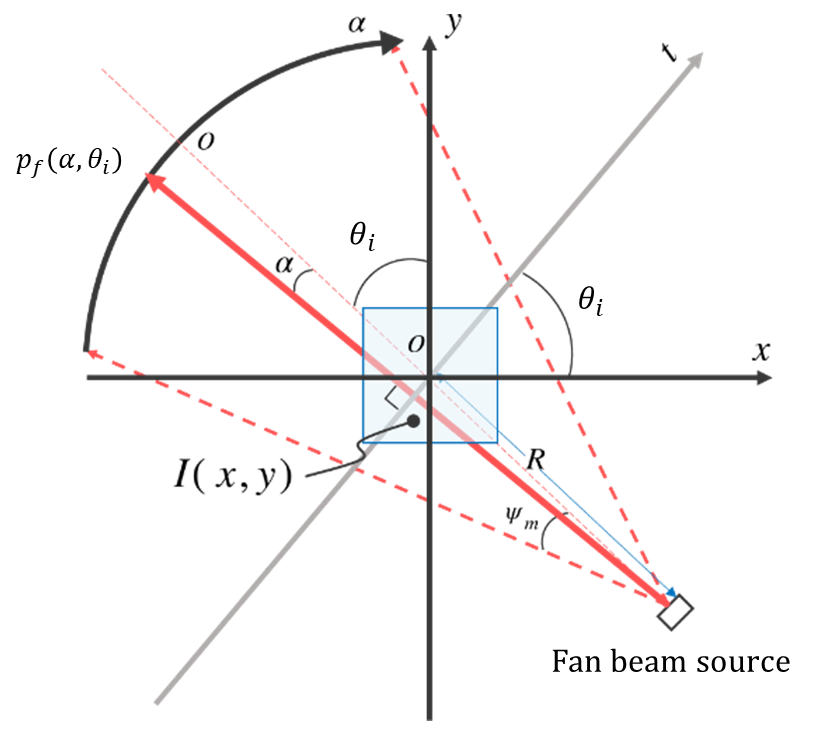}
  \parbox[t]{.9\columnwidth}{\relax
  }
  \caption{\label{fig:radon}
  CT-based aerial image formation. Radon transform computes projection data from the target image, which is then converted to fan-beam data. The projection images are projected from multiple angles and superimposed in fog to form an aerial image.
           }
\end{figure}

Since projector rays diverge radially, back-projection with projector rays can be formulated as fan-beam CT (or cone-beam CT in the case of three-dimensional projection) (Fig.~\ref{fig:radon}). For computational efficiency, however, this study first computes projection data using parallel-beam Radon transform and then converts it into fan-beam projection data.

For a target image $I(x,y)$, the projection data $p(t,\theta)$ at projection angle $\theta$ is obtained by the Radon transform:

\begin{equation}
p(t,\theta) = \int\!\!\int I(x,y)\,\delta(x\cos\theta + y\sin\theta - t)\, dx\, dy ,
\label{eq:radon}
\end{equation}

\noindent
where $\delta(\cdot)$ denotes the Dirac delta function and $t$ is the offset of the projection line.
Practically, Eq.~\ref{eq:radon} can be computed by rotating the image by angle $\theta$ and then summing the pixel values along the vertical (or horizontal) axis.

\subsection{Fan-beam conversion}

To account for the divergence of projector rays, the parallel-beam data $p(t,\theta_i)$ obtained by the Radon transform is converted into fan-beam data $p_f(\alpha,\theta_i)$:

\begin{equation}
p_f(\alpha, \theta_i) = p(R \sin\alpha, \theta_i + \alpha),
\label{eq:fanbeam}
\end{equation}

\noindent
where $\theta_i$ is the placement angle of the $i$-th projector, $\alpha$ is the angular offset of each ray, and $R$ is the placement radius.

\subsection{Generation of projection images}

Based on Eq.~\ref{eq:fanbeam}, the projection data are mapped to grayscale values to generate the projection image for each projector. To display a horizontal-plane image in mid-air, the projection image $I_i(\alpha,\theta_i)$ for projector $i$ was created by assigning the projection data $p_f(\alpha,\theta_i)$ to a horizontal band of 30 pixels centered in the image, while filling the other pixels with black. The projection data were normalized and mapped to intensity values in the range 0-255.

\subsection{Back-projection for aerial image formation}

The generated projection images are projected into fog from multiple projectors, and the rays are superimposed to form an aerial image. The reconstructed image by simple back-projection is expressed as

\begin{equation}
I'(x,y) = \frac{1}{N} \sum_{i=0}^{N-1} p_f(\alpha, \theta_i),
\label{eq:backproj}
\end{equation}

\noindent
where $N$ is the number of projectors. Theoretically, if $N$ is sufficiently large, back-projection can approximate an arbitrary image with blurring. Therefore, the proposed method is in principle capable of representing textures and grayscale levels, which have been difficult to realize with previous approaches.

\section{Experiment}

To evaluate the effectiveness of the proposed method, we conducted both simulations and preliminary experiments in a real environment.

\subsection{Apparatus}

\begin{figure}[tb]
  \centering
  \includegraphics[width=.8\linewidth]{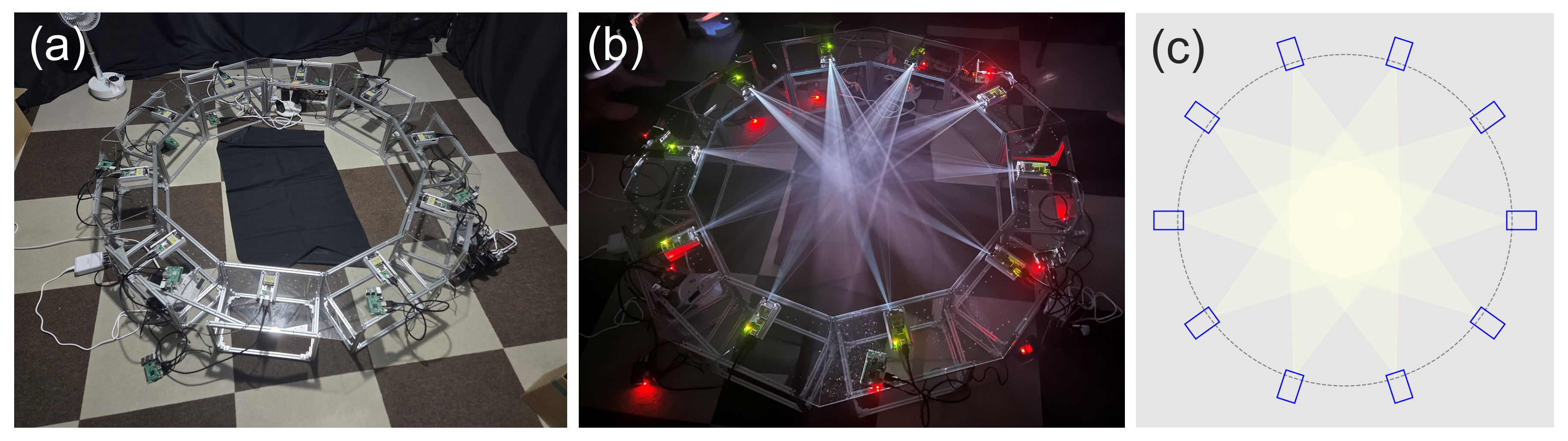}
  \parbox[t]{.9\columnwidth}{\relax
  }
  \caption{\label{fig:apparatus}
           CT-based volumetric display. (a) External view. The projectors are arranged on a circular platform, and the rays are projected into the fog. (b) External view while projecting. The projector rays were visualized by fog scattering. (c) Arrangement and projection area. The projectors are oriented toward the center of the platform, with angular intervals of 36\textdegree{}.
           }
\end{figure}

An overview of the projection apparatus is shown in Fig.~\ref{fig:apparatus}a. We used a compact scanning laser projector (HD309D1-C1, Ultimems Inc., resolution 1280 $\times$ 720 pixels, horizontal field of view 39\textdegree{}) for projection. Each projector was controlled by a small computer (Raspberry Pi 3 Model B v1.2) connected via HDMI. Projection images were generated on a main PC and transmitted to each small computer via Wi-Fi. Since only static images were projected in the experiments, the images were preloaded, and all projectors were triggered simultaneously by the main PC via a Wi-Fi command, ensuring frame-level synchronization.

Accurate alignment of projectors is critical, because misalignment directly results in errors in the position of luminous points. To ensure precise placement, we designed a circular mounting stage using acrylic plates and aluminum frames to fix the projectors. Each projector was fine-tuned by projecting a cross pattern and aligning all vertical lines to intersect at the center of the apparatus. The height was also adjusted to avoid reflections from the table surface interfering with the display volume.

To visualize the projected rays in air, a fog machine (SM400 Tapir, Sound House Co.) with a water-based fog liquid (FLX4) was employed. A particle sensor (sps30, Sensirion) measured the fog particle distribution as approximately 3:12:7:1 for PM0.5, PM1.0, PM2.0, and PM4.0, respectively. Experiments were conducted in a 3 m $\times$ 3 m darkroom booth with external light blocked. It was observed that the projector rays were well visualized by Mie scattering (Fig. ~\ref{fig:apparatus}b).

The projector arrangement is illustrated in Fig.~\ref{fig:apparatus}c. The projectors were placed on a circular stage with a radius of 0.6 m and oriented toward the center. In the prototype reported here, the number of projectors was set to $N=10$, evenly distributed in angle. This configuration represents a preliminary stage toward a larger setup with $N=50$ projectors \cite{Yamada25-IWAIT}, which is expected to improve image quality.

\subsection{Results}

\begin{figure}[tb]
  \centering
  \includegraphics[width=.8\linewidth]{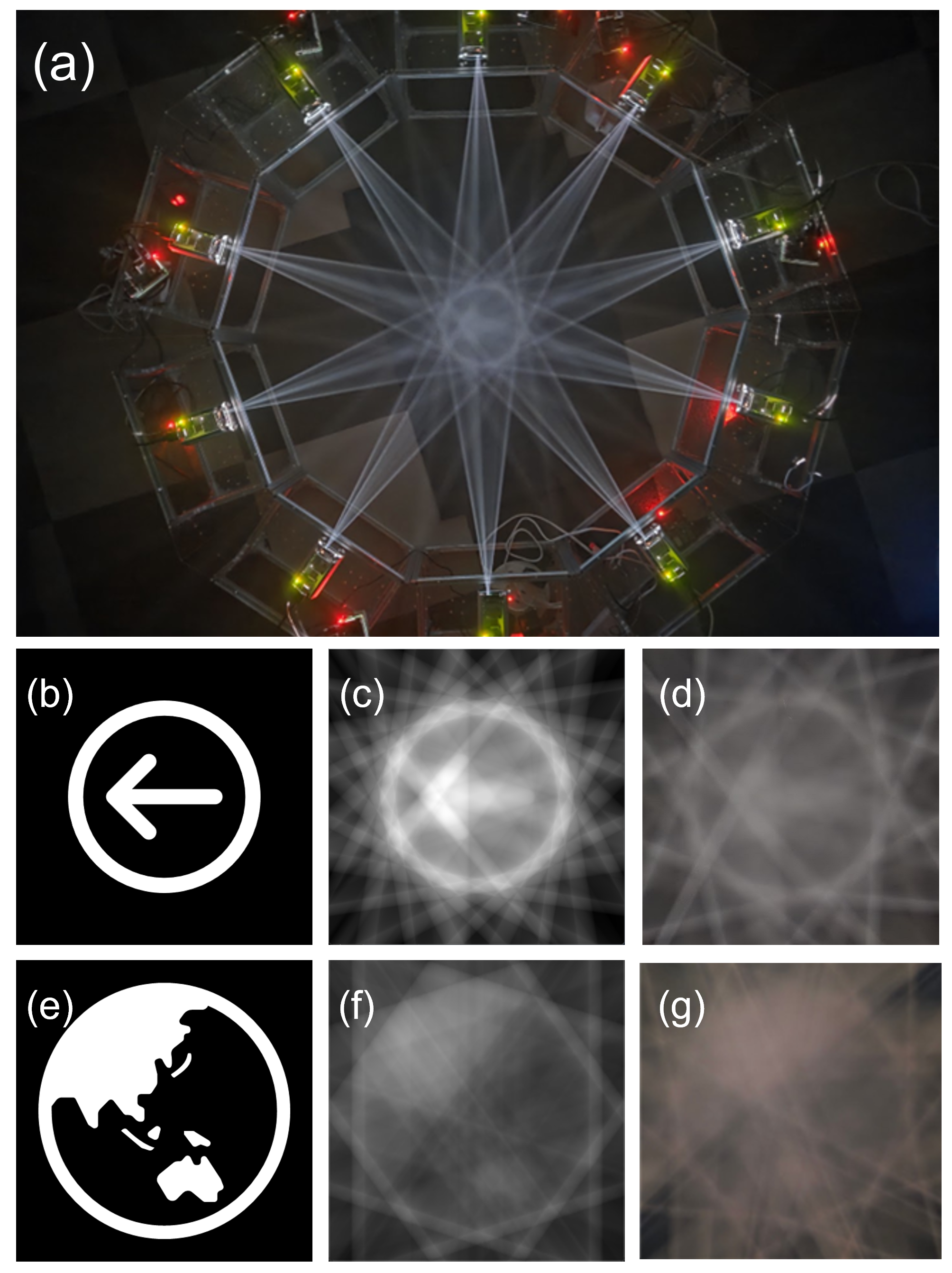}
  \parbox[t]{.9\columnwidth}{\relax
  }
  \caption{\label{fig:results}
           The results of 2D aerial images formed by CT-based back-projection with $N=10$ projectors. (a) External top view of aerial image formation. (b,e) Target images (simple and complex). (c,f) Simulation results generated under identical parameters (simple back-projection). (d,g) Photographs of the aerial images taken from above by smartphone. The images are faint but recognizable, demonstrating the feasibility of the proposed method.
           }
\end{figure}

Figure~\ref{fig:results} shows the projection results obtained in a real environment. As illustrated in Fig.~\ref{fig:results}a, projector beams visualized in fog converge near the center of the circular setup, forming an aerial image observable from above. Two types of target images were tested: simple shapes with hollow regions (an arrow, Fig.~\ref{fig:results}b--d) and more complex patterns (a globe icon, Fig.~\ref{fig:results}e--g). The reconstructed results (Fig.~\ref{fig:results}d,g) closely match the simulations with the same number of projectors ($N=10$, Fig.~\ref{fig:results}c,f), indicating that the intended aerial images were successfully reproduced. In both cases, the overall outlines of the shapes were recognizable, and in particular, the direction of the arrow could be clearly identified (Fig.~\ref{fig:results}d). However, consistent with the simulation, fine details inside the complex shape (Fig.~\ref{fig:results}e) were difficult to perceive.

\section{Discussion}

The experiments confirmed that images generated in simulation were reproduced as aerial images in practice. Even at the present time, with a small number of projectors, simple signs (arrow directions) were still visible. Conventional methods cannot display hollow structures on a horizontal plane, indicating that the proposed CT-based approach is valid even in a real environment.

The most significant distinction of the proposed method from prior work such as RayGraphy \cite{Yamada20-RayGraphy} lies in its tomographic formulation. In RayGraphy, each projector simply casts a silhouette of the object from its own viewpoint; thus, the resulting image is limited to binary contours and cannot, in principle, represent cross-sectional slices on a horizontal plane or internal gradations. Therefore, the aerial 2D cross-sectional images demonstrated in our experiment (Fig. \ref{fig:results}d,g) could not have been achieved by such silhouette-based methods.
In contrast, our approach treats aerial image formation as a CT-style back-projection problem, enabling the reconstruction of arbitrary voxel distributions --- including concave regions, textures, and color or brightness variations --- as long as a sufficient number of projection angles are available.
While the present experiment demonstrates only two-dimensional aerial images, the same principle directly extends to full three-dimensional volumetric projection, distinguishing it fundamentally from fog-screen projection that merely displays images on a surface of mist.

The generated images, however, appeared blurred and lacked fine details. This outcome was consistent with simulation and can be attributed to two main factors. First, the number of projectors was insufficient; prior work \cite{Yamada25-IWAIT} suggests that increasing the number of projectors can improve image quality. Second, the present results relied on simple back-projection. In CT, image reconstruction using the simple back-projection method is known to suffer from a lack of high-frequency components and overall blurring of the image. Introducing filtered back-projection (FBP), which is widely used in CT reconstruction, may alleviate this problem. A challenge arises because filtered projection data can include negative values that cannot be directly represented by projector light. One possible solution is to normalize the data and map negative values to gradations below a baseline level, though this may reduce image contrast. We plan to investigate this through both simulation and further experiments.

In addition, the aerial images were observable only from the top view, while near-horizontal viewing was difficult. This limitation is likely due to the anisotropic nature of Mie scattering, by which light is scattered more strongly in directions close to its original path. As a result, opposing projector rays were more strongly scattered in horizontal directions, preventing accurate ray superposition and degrading the image.
Other differences between simulated and experimental images were also observed. These may be attributed to anisotropic scattering, slight deviations in projector alignment, light attenuation, and non-uniformity in fog particle distribution. To address these issues, future work should incorporate simulations that explicitly model such optical effects.

At this stage, our evaluation was limited to qualitative observations. To better understand these differences, quantitative measures such as brightness, contrast, and perceptual recognition rates will also be investigated.

Finally, we briefly mention the potential user experience that this method could offer in the future. At the current stage, our work remains at the level of simulations and a proof-of-concept with a small number of projectors, and no user study has yet been conducted. However, because the proposed method does not require a physical screen in the display volume, the system could in principle be installed in indoor spaces ranging from small showrooms to large concert halls. Moreover, except along the projector directions, multiple viewers would be able to observe the same aerial image simultaneously from arbitrary viewpoints. Such a shared aerial image could be particularly useful in scenarios such as education, collaborative work, and presentations. Furthermore, as shown in this study, simple signs can be displayed in the air that can be passed through, which could be applied to traffic signs in tunnels. In addition, since the system generates volumetric images as actual luminous points suspended in free space and is, in principle, capable of high frame rate display, interactive applications in which users directly reach out to the images and receive responses can be considered as future research topics. At present, the images are faint and require a darkened environment, but with an increased number of projectors and improved reconstruction algorithms, evaluations involving actual user experiences will become feasible in the future.

\section{Conclusion and Future Work}

We proposed a ray-superimposed volumetric display based on CT back-projection and conducted preliminary investigations. As an initial feasibility study, a prototype with ten projectors successfully demonstrated the display of two-dimensional figures in mid-air.

In the future, we will increase the number of projectors, introduce filtered back-projection to reduce blurring, and establish calibration methods to compensate for alignment errors. We will also extend the method to three-dimensional aerial images and explore applications such as in education, architecture, and entertainment.

\section*{Acknowledgments}
This work was supported by JSPS KAKENHI Grant Number 25K21252.

\input{main.bbl}

\end{document}

%% file: main.bbl
\newcommand{\etalchar}[1]{$^{#1}$}